# Reproducibility in medical image radiomic studies: contribution of dynamic histogram binning

Darryl E. Wright, Ph.D.
Mayo Clinic, Radiology, Rochester, MN, USA

Cole Cook, Ph.D., M.S.
Mayo Clinic, Radiology, Rochester, MN, USA

Jason Klug, Ph.D.
Mayo Clinic, Radiology, Rochester, MN, USA

Panagiotis Korfiatis, Ph.D.
Mayo Clinic, Radiology, Rochester, MN, USA

Timothy L. Kline, Ph.D.
Mayo Clinic, Radiology, Rochester, MN, USA
kline.timothy@mayo.edu
(507)-538-2164

Mayo Clinic, 200 1st St SW, Rochester, MN 55905

Research Letter

Key results:

- Dynamic histogram binning is sensitive to small changes in the region from which radiomic features are extracted and may contribute to reproducibility concerns for radiomic-based machine learning.

Abbreviations:

- NGLDM - neighboring gray-level dependence matrix
- LDLGLE - low dependence low gray-level emphasis
- LDE - low dependence emphasis
- CCC - concordance correlation coefficient

## Introduction

The de facto standard of dynamic histogram binning for radiomic feature extraction leads to an elevated sensitivity to fluctuations in annotated regions. This may impact the majority of radiomic studies published recently and contribute to issues regarding poor reproducibility of radiomic-based machine learning [1] that has led to significant efforts for data harmonization [2]; however, we believe the issues highlighted here are comparatively neglected, but often remedied by choosing static binning.

The field of radiomics has improved through the development of community standards [3] and open-source libraries such as PyRadiomics [4]. But differences in image acquisition, systematic differences between observers' annotations, and preprocessing steps still pose challenges. These can change the distribution of voxels altering extracted features and can be exacerbated with dynamic binning.

## Materials and Methods

To illustrate, we use screening mammography obtained at Mayo Clinic between January 2009 and December 2019 comprising 4095 images with lesions confirmed as malignant (n=2005) or benign (n=2090) by biopsy. Bounding boxes locating the lesions were drawn by radiologists. All subjects assented to research authorization in accordance with the Mayo Clinic Institutional Review Board.

Images and masks were resampled to the median pixel dimensions of 0.07mm x 0.07mm using b-spline and nearest neighbor interpolation, respectively. Images were further normalized to the interval [0, 255]. Radiomic features were extracted from the preprocessed images using either PyRadiomics version 3.0.1, where dynamic binning was performed or a modified PyRadiomics code implementing static binning.

To investigate feature robustness, the original bounding boxes were eroded or dilated until the area was altered by 20%, simulating feature robustness to inter-reader annotations. Two features from the neighboring gray-level dependence matrix (NGLDM) were calculated from these images: (1) the low dependence low gray-level emphasis (LDLGLE) and (2) the low dependence emphasis (LDE). These features were chosen since LDE is independent of the gray-level of each element of the NGLDM while the LDLGLE is not.

Similarity metrics were computed for the radiomic features between the 20% eroded and original bounding boxes using dynamic and static binning. The metrics were: Pearson correlation, Spearman correlation, and Lin's concordance correlation coefficient.

For each image, the feature value derived from the original bounding box was plotted against that derived from the 20% eroded version. A feature robust to morphological operations would fall close to y=x.

## Results

Table 1(2) shows summary statistics for the similarity between the radiomic features extracted from the 20% eroded and 20% dilated bounding boxes for the dynamic(static) binning. Metrics were consistently higher with static than dynamic binning and more features showed agreement >0.9.

Figure 1 shows the effect of eroding the bounding boxes by 20% compared to the original for (a) LDLGLE and (b) LDE each extracted with static and dynamic binning. In both panels (a) and (b), the magenta and yellow points denote the locations of the images shown in panels (c) and (d), respectively.

Many points for LDLGLE show good agreement. However, for dynamic binning a significant minority of points are less correlated. For LDE, all points show good agreement irrespective of the binning, as expected. Panel (c) corresponds to an image where the radiomic features are robust to erosion whatever the binning. Panel (d) shows an image that lies off y=x and where the feature is not robust.

## Discussion

The better agreement between features with static than with dynamic binning demonstrates the robustness issue we wish to highlight. Defaulting to dynamic binning (as in PyRadiomics) might lead to unexplained or misattributed robustness issues leading to results that are difficult to replicate. Whole image binning and/or study specific ranges should become standard practice for reproducibility of machine learning models trained on image texture features.

Table 1: Summary statistics measuring the similarity between the radiomic features extracted from the 20% eroded and 20% dilated bounding boxes for the dynamic binning case.

| Metric | Minimum | Median | Mean | Maximum | Standard Deviation | n > 0.9 (86 total) |
|---|---|---|---|---|---|---|
| **Pearson** | 0.829 | 0.977 | 0.970 | 0.999 | 0.032 | 82 |
| **Spearman** | 0.883 | 0.981 | 0.973 | 0.999 | 0.026 | 84 |
| **Lin's CCC** | 0.764 | 0.967 | 0.953 | 0.997 | 0.044 | 75 |

Table 2: Summary statistics measuring the similarity between the radiomic features extracted from the 20% eroded and 20% dilated bounding boxes for the static binning case.

| Metric | Minimum | Median | Mean | Maximum | Standard Deviation | n > 0.9 (86 total) |
|---|---|---|---|---|---|---|
| **Pearson** | 0.902 | 0.985 | 0.981 | 0.999 | 0.019 | 86 |
| **Spearman** | 0.914 | 0.985 | 0.982 | 0.999 | 0.017 | 86 |
| **Lin's CCC** | 0.856 | 0.971 | 0.962 | 0.997 | 0.033 | 80 |

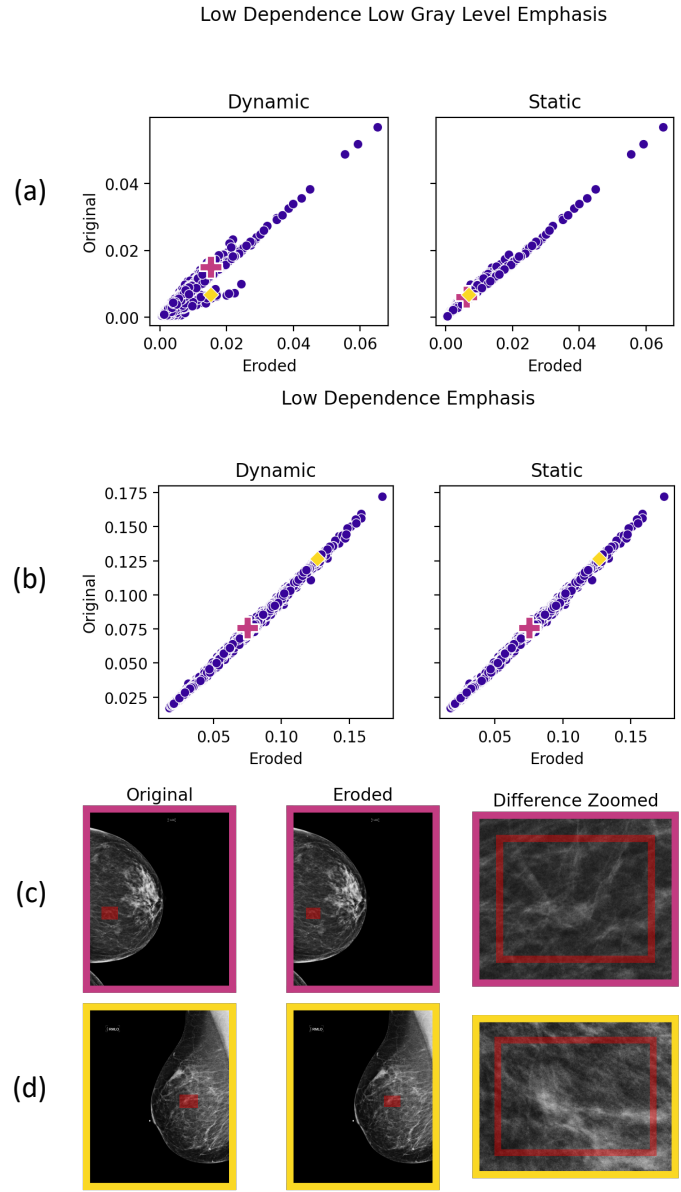

Figure 2: (a) Comparison of low dependence low gray level emphasis extracted from the original and 20% eroded bounding boxes using dynamic and static binning. (b) The same for low dependence emphasis. (c) and (d) The first and second columns show the original and 20% eroded bounding boxes overlaid for the full images corresponding to the magenta and yellow points in panels (a) and (b). The third column shows only the image region contained by the original bounding box with the 20% eroded version in red for comparison.